# Optical Signatures of Spin-Orbit Exciton in Bandwidth Controlled $Sr_2IrO_4$ Epitaxial Films via High-Concentration Ca and Ba Doping

M. Souri,[1] B. H. Kim,[2,3] J. H. Gruenewald,[1] J. G. Connell,[1] J. Thompson,[1] J. Nichols,[1] J. Terzic,[1,4] B. I. Min,[2] G. Cao,[1,4] J. W. Brill,[1] and A. Seo[1,*]

[1] Department of Physics and Astronomy, University of Kentucky, Lexington, KY 40506, USA
[2] Department of Physics, PCTP, Pohang University of Science and Technology, Pohang 790-784, Korea
[3] iTHES Research Group and Computational Condensed Matter Physics Laboratory, RIKEN, Wako, Saitama 351- 0198, Japan
[4] Department of Physics, University of Colorado at Boulder, Boulder, CO 80309, USA

## Abstract

We have investigated the electronic and optical properties of $(Sr_{1-x}Ca_x)_2IrO_4$ ($x = 0 - 0.375$) and $(Sr_{1-y}Ba_y)_2IrO_4$ ($y = 0 - 0.375$) epitaxial thin-films, in which the bandwidth is systematically tuned via chemical substitutions of Sr ions by Ca and Ba. Transport measurements indicate that the thin-film series exhibits insulating behavior, similar to the $J_{eff} = 1/2$ spin-orbit Mott insulator $Sr_2IrO_4$. As the average A-site ionic radius increases from $(Sr_{1-x}Ca_x)_2IrO_4$ to $(Sr_{1-y}Ba_y)_2IrO_4$, optical conductivity spectra in the near-infrared region shift to lower energies, which cannot be explained by the simple picture of well-separated $J_{eff} = 1/2$ and $J_{eff} = 3/2$ bands. We suggest that the two-peak-like optical conductivity spectra of the layered iridates originates from the overlap between the optically-forbidden spin-orbit exciton and the inter-site optical transitions within the $J_{eff} = 1/2$ band. Our experimental results are consistent with this interpretation as implemented by a multi-orbital Hubbard model calculation: namely, incorporating a strong Fano-like coupling between the spin-orbit exciton and inter-site $d$-$d$ transitions within the $J_{eff} = 1/2$ band.

I.     INTRODUCTION

Complex iridium oxides (iridates) such as the Ruddlesden-Popper (RP) series $Sr_{n+1}Ir_nO_{3n+1}$ ($n = 1 - \infty$) have attracted substantial attention due to their novel electronic states originating from coexisting strong spin-orbit interaction and electron-correlation. The strong spin-orbit interaction in iridium splits the $t_{2g}$ band of $5d^5$ electrons into the fully-occupied $J_{eff} = 3/2$ and half-filled $J_{eff} = 1/2$ band. When the bandwidth of the half-filled $J_{eff} = 1/2$ band is large enough (e.g. $SrIrO_3$), the system exhibits strongly correlated metallic behavior [1, 2]. However, as the dimensionality of the system decreases, a Mott insulating gap is opened due to reduced bandwidth, resulting in a $J_{eff} = 1/2$ Mott insulator (e.g. $Sr_2IrO_4$), which has been identified by angled resolved photoemission spectroscopy, x-ray absorption spectroscopy, optical spectroscopy, and first-principle calculations [3-5]. The $J_{eff} = 1/2$ Mott insulator $Sr_2IrO_4$ has been theoretically suggested as a prospective compound for new high-$T_c$ superconducting states [6, 7], and $d$-wave gap symmetry has been experimentally observed with electron doping on its surface [8-10]. While layered iridates are attracting substantial attention, there remain controversial issues regarding the fundamental electronic structure of this system: 1) The origin of the insulating gap is disputed as arising either from an antiferromagnetic ordering, i.e. Slater scheme [11, 12], or electron-correlation, i.e. Mott scheme [3, 4]. 2) Evidence for the strong hybridization of the $J_{eff} = 1/2$ and $J_{eff} = 3/2$ states [11, 13] is incompatible with the conventional picture of well-separated $J_{eff} = 1/2$ and $J_{eff} = 3/2$ bands.

In this paper, we report the effects of tuning the bandwidth via chemical pressure (i.e. Ca and Ba doping) on the electronic and optical properties of $Sr_2IrO_4$ epitaxial thin-films. Bandwidth control of the $J_{eff} = 1/2$ state has been employed to better understand the metal-insulator transition and the electronic structure of RP series iridates with different dimensionalities [1]. Here, we focus on using different ionic sizes of Ca ($r_{Ca^{2+}} = 1.14$ Å), Ba ($r_{Ba^{2+}} = 1.49$ Å), and Sr ($r_{Sr^{2+}} = 1.32$ Å) ions [14-16]: substitution of Sr by Ca and Ba ions exerts chemical pressure without changing the 4+ valence state of the Ir ions. Due to the smaller (larger) ionic size of $Ca^{2+}$ ($Ba^{2+}$) compared to $Sr^{2+}$, Ca (Ba) doping in $Sr_2IrO_4$ decreases (increases) the in-plane Ir-O-Ir bond angle ($\theta$). According to the relation between the bandwidth ($W$) and $\theta$:

$$W \approx \frac{\cos\frac{\pi-\theta}{2}}{d_{Ir-O}^{3.5}} \qquad (1)$$

where $d_{Ir-O}$ is the Ir-O bond length, Ca (Ba) doping decreases (increases) both $W$ and the electronic hopping integral ($t$) [17]. The decreased (increased) $W$ affects the effective electron-correlation energy, $U_{eff} \equiv U/W$, where $U$ is the on-site Coulomb repulsion. To explore the effects of the decreased (increased) $W$ on the optical and electronic properties, we have synthesized $K_2NiF_4$-type tetragonal $(Sr_{1-x}Ca_x)_2IrO_4$ and $(Sr_{1-y}Ba_y)_2IrO_4$ epitaxial thin-films with $x$ and $y = 0.125$ (1/8), 0.25 (1/4), and 0.375 (3/8), by epitaxial stabilization. Note that these high doping concentrations of Ca and Ba ions beyond the solubility limit are not readily achievable by conventional solid state chemistry since the tetragonal phase bulk crystals of $(Sr_{1-x}Ca_x)_2IrO_4$ and $(Sr_{1-y}Ba_y)_2IrO_4$ are stable only for ($x = 0 - 0.11$) and ($y = 0 - 0.1$) [18, 19]. Epitaxial compressive (tensile) strain can also increase (decrease) the in-plane $IrO_6$ octahedral rotation and decreases (increases) $W$ in the system [20]. However, high-concentration doping with smaller (larger) A-site ions increases (decreases) both the $IrO_6$ octahedral rotation and tilting in the system, which tunes the electronic structure effectively. In order to consider only the effect of A-site doping in the system, we have grown all of the $(Sr_{1-x}A_x)_2IrO_4$ (A: Ca, Ba) thin-films on the same ($SrTiO_3$ (100) (STO)) substrates. From optical spectroscopic characterizations, we have observed an unexpected shift in the optical conductivity spectra to lower energies as $W$ increases from ($Sr_1$- $_x$Ca$_x$)$_2$IrO$_4$ to (Sr$_{1-y}$Ba$_y$)$_2$IrO$_4$. This red-shift in $\sigma_1(\omega)$ cannot be explained by the simple picture of well-separated $J_{\text{eff}} = 1/2$ and $J_{\text{eff}} = 3/2$ bands [3-5]. Using multi-orbital Hubbard model calculations, we propose that the overall $\sigma_1(\omega)$ spectral shape of the layered iridates originates from Fano-like coupling between inter-site *d-d* transitions within the $J_{\text{eff}} = 1/2$ band and the optically-forbidden spin-orbit exciton—which correctly shows the red-shift in $\sigma_1(\omega)$ as $U_{\text{eff}}$ decreases in this system.

## II. METHODS

Epitaxial thin-films of (Sr$_{1-x}$Ca$_x$)$_2$IrO$_4$ and (Sr$_{1-y}$Ba$_y$)$_2$IrO$_4$ ($x$ and $y = 0.125, 0.25, 0.375$) with the K$_2$NiF$_4$–type structure have been synthesized by pulsed laser deposition (PLD). The thin-films are grown on STO substrates with a laser fluence of 1.2 J/cm$^2$ (KrF excimer, $\lambda = 248$ nm), a substrate temperature of 700 °C, and 10 mTorr oxygen partial pressure by alternating a Sr$_2$IrO$_4$ (*I4$_1$/acd*) target, a Ca$_2$IrO$_4$ (*P62m*) target, and a ceramic target with Ba:Ir = 2:1 stoichiometry comprised mostly of the BaIrO$_3$ phase (*C2/m*) and BaO [21, 22]. Atomically flat TiO$_2$-terminated STO substrates are prepared using the method described in Ref. [23]. In order to stabilize the K$_2$NiF$_4$-type tetragonal structure, we have used the technique of controlling PLD plume dimensions as reported in Ref. [24]. We have checked the Ba and Ca concentrations of our samples via energy dispersive x-ray (EDX) spectroscopy. The average concentrations of Ba and Ca ions are found to be within ± 3% of the nominal values of $x$ and $y = 0.125, 0.25, 0.375$ in (Sr$_{1-x}$Ca$_x$)$_2$IrO$_4$ and (Sr$_{1-y}$Ba$_y$)$_2$IrO$_4$ thin-films. The epitaxial K$_2$NiF$_4$-type structure of our thin-films has been confirmed using x-ray diffraction. The transport properties have been measured by using a conventional four-probe method. The optical transmittance spectra of the thin-films have been taken at normal incidence using a Fourier-transform infrared spectrometer and a grating-type spectrophotometer in the photon energy regions of 0.06 – 0.5 eV and 0.5 – 3.2 eV, respectively. Due to the Reststrahlen band of STO substrates, 0.2 eV is the lowest photon energy limit for the transmittance spectra. We have obtained the in-plane optical conductivity spectra ($\sigma_1(\omega)$) using the Kramers-Kronig transformation. We have numerically calculated the excitation and $\sigma_1(\omega)$ as a function of the electronic hopping integral, which is proportional to the bandwidth ($W$), by solving a multi-orbital Hubbard model [25] including the on-site Coulomb interaction between 5$d$ electrons ($U = 1.86$ eV) and the spin-orbit coupling of iridates ($\lambda_{SO} = 0.48$ eV).

### III. EXPERIMENTAL RESULTS

Figure 1 (a) shows the $\theta$-$2\theta$ x-ray diffraction scans confirming the $c$-axis orientation of $(Sr_{1-x}Ca_x)_2IrO_4$ and $(Sr_{1-y}Ba_y)_2IrO_4$ thin-films. The enlarged scans in Fig. 1 (b) clearly show that the (00$\underline{12}$) reflections of the thin-films are shifted to lower angles as the out-of-plane lattice parameters become larger from $(Sr_{1-x}Ca_x)_2IrO_4$ to $(Sr_{1-y}Ba_y)_2IrO_4$. The thickness of the thin films is ca. 20 nm. X-ray reciprocal space mapping around the (103) reflection of STO (Fig. 2 (a)) shows the (11$\underline{18}$) reflections of the thin-films. The vertical alignment of the thin-film peak with that of the substrate indicates that the films are coherently strained in plane. Figure 2 (b) summarizes the in-plane ($a$) and out-of-plane ($c$) lattice parameters as a function of average A-site ionic radius in $(Sr_{1-x}Ca_x)_2IrO_4$ and $(Sr_{1-y}Ba_y)_2IrO_4$ thin-films. While the in-plane lattice parameters are constant, the out-of-plane lattice parameters increase systematically as the average A-site ionic radius increases. The temperature dependence of the resistivity reveals that all of the $(Sr_{1-x}Ca_x)_2IrO_4$ and $(Sr_{1-y}Ba_y)_2IrO_4$ films exhibit insulating behavior (Fig. 2 (c)). The room-temperature resistivity of the samples is in the range of 100 - 400 m$\Omega$·cm. The decreased $U_{eff}$ from $(Sr_{1-x}Ca_x)_2IrO_4$ to $(Sr_{1-y}Ba_y)_2IrO_4$ would be expected to systematically decrease the resistivity. However, all of the doped samples have lower resistivity than the pure Sr$_2$IrO$_4$ thin-film, which implies that the transport properties of the doped layered iridates are dominated by impurities or defects. Various impurities and defects such as oxygen vacancies may increase the carrier concentrations of the samples by doping electrons while the samples remain insulating. Note also that tetragonal Ca$_2$IrO$_4$ and Ba$_2$IrO$_4$ are thermodynamically metastable phases; hence, we have used a Ca$_2$IrO$_4$ target with a hexagonal structure and a Ba$_2$IrO$_4$ target comprised mostly of the BaIrO$_3$ phase, which may further increase the amount of unidentified impurities or dopants.

Figure 3 shows $\sigma_1(\omega)$ of (Sr$_{1-x}$Ca$_x$)$_2$IrO$_4$ and (Sr$_{1-y}$Ba$_y$)$_2$IrO$_4$ thin-films with the Lorentz oscillator fits. In order to have the best fit with experimental spectra, we need a minimum of three or four oscillators in the 0.2 – 2 eV photon energy region, which is shown with the fit curves of thin green, red, blue, and orange colors. The thin green oscillator, located at ~ 0.25 eV, has been indicated as the inner-gap excitation peak [26]. The high energy tails of the charge-transfer transitions from O 2$p$ band to Ir 5$d$ band above 3 eV and the weak optical transitions from the Ir 5$d$ $t_{2g}$ band to the Ir 5$d$ $e_g$ band above 2 eV are both shown by thin gray curves. The black curves show the total spectra of the fit, which matches well with the experimental spectra. Note that, as it is illustrated in the Fig. 3, there are two-peak features (the so-called $\alpha$ peak (thin red curve) at ~ 0.5 eV and $\beta$ peak (thin blue curve) at ~ 0.8 eV) in $\sigma_1(\omega)$, which have been interpreted as two separate optical transitions from the $J_{\text{eff}} = 1/2$ lower Hubbard band ($\alpha$) to the $J_{\text{eff}} = 1/2$ upper Hubbard band and the $J_{\text{eff}} = 3/2$ band ($\beta$) to the $J_{\text{eff}} = 1/2$ upper Hubbard band in Sr$_2$IrO$_4$ [3-5]. Our optical results show that as the average A-site radius increases from (Sr$_{1-x}$Ca$_x$)$_2$IrO$_4$ to (Sr$_{1-y}$Ba$_y$)$_2$IrO$_4$, both $\alpha$ and $\beta$ peak positions *shift* to lower energies (Fig. 4 (b)) and the spectral weight ratio between $\alpha$ and $\beta$ transitions (SW$_\alpha$/SW$_\beta$) also increases (Fig. 4 (b)). According to the simple picture of well-separated $J_{\text{eff}} = 1/2$ and $J_{\text{eff}} = 3/2$ bands, the decreased $U_{\text{eff}}$ from (Sr$_{1-x}$Ca$_x$)$_2$IrO$_4$ to $(Sr_{1-y}Ba_y)_2IrO_4$ results in a decreased separation between the lower Hubbard band (LHB) and upper Hubbard band (UHB). The decrease in the separation causes a shift in the $\alpha$ peak position to lower energy, which is consistent with the red-shift in the $\alpha$ peak position in our optical results (Figs. 3 and 4 (a)). However, understanding the shift in the $\beta$ peak position of our optical data using this picture is challenging. In order to explain the shift in the $\beta$ peak position based on this picture we consider three different scenarios: 1) As $U_{eff}$ decreases the separation between the LHB and UHB should decrease. Hence, both LHB and UHB should shift. Since the $\alpha$ peak position is the transition from the $J_{eff} = 1/2$ LHB to the $J_{eff} = 1/2$ UHB, the shift in the $\alpha$ peak position is dependent on the shift in both bands. However, the shift in the $\beta$ peak position, which is the transition from the $J_{eff} = 3/2$ band to the $J_{eff} = 1/2$ UHB, only depends on the UHB shift. This scenario indicates that the shift in the $\beta$ peak position should be approximately half of the $\alpha$ peak position shift [1, 27], which is not consistent with the observed shift in the $\beta$ peak in our optical data (Fig. 3 and Fig. 4 (a)). 2) If we assume the LHB to be fixed and only the UHB to shift to lower energy, the shift in the $\beta$ peak position should be equal to the shift in the $\alpha$ peak position, which is consistent with our optical data. However, this picture cannot explain the observed changes in the spectral weights between $\alpha$ and $\beta$ transitions (Fig. 4 (b)). 3) Since the peak position is determined by the separation between the $J_{eff} = 3/2$ band and the $J_{eff} = 1/2$ UHB, the $\beta$ peak energy should be proportional to the spin-orbit coupling energy ($\lambda_{SO}$) [1, 5]. Since $\lambda_{SO}$ is determined by the atomic number of the iridium (Z), i.e. $\lambda_{SO} \propto Z^4$, it should remain constant in all $(Sr_{1-x}A_x)_2IrO_4$ thin-films, which does not explain the shift in the $\beta$ peak in our optical data (Fig. 3). Hence, the simple picture of well-separated $J_{eff} = 1/2$ and $J_{eff} = 3/2$ bands cannot explain $\sigma_1(\omega)$ shifting to lower energy region. Moreover, in this picture with well-separated $J_{eff} = 1/2$ and $J_{eff} = 3/2$ bands, the spectral weight ratio ($SW_\alpha/SW_\beta$) should be constant, which does not explain the increased spectral weight ratio ($SW_\alpha/SW_\beta$) between $\alpha$ and $\beta$ optical transitions (Fig. 4 (b)) from $(Sr_{1-x}Ca_x)_2IrO_4$ to $(Sr_{1-y}Ba_y)_2IrO_4$. Hence, our experimental observations call the current interpretation of $\sigma_1(\omega)$ for layered iridates based on well-separated $J_{eff} = 1/2$ and $J_{eff} = 3/2$ bands into question. Recent theoretical studies also indicate that strong hybridization of the $J_{eff} = 1/2$ and 3/2 states causes a mixing of their energies [13].

## IV. DISCUSSION

We suggest that the optically-forbidden spin-orbit exciton overlaps with an electronic continuum originating from the inter-site *d-d* transitions within the $J_{eff} = 1/2$ band, which results in the two-peak-like structures observed in $\sigma_1(\omega)$ of layered iridates. Resonant inelastic x-ray scattering (RIXS) experiments on $Sr_2IrO_4$ crystals have discovered charge-neutral excitations, which are referred to as the spin-orbit exciton or *spin-orbiton* [25, 28, 29]. These neutral particles originate from intra-site electron-hole pairs, i.e. a hole in the $J_{eff} = 3/2$ band and an electron in $J_{eff} = 1/2$ band, that move through the lattice and create a tail of flipped spins in the ground state of the system. The energy of this exciton is similar to the energy of spin-orbit coupling as observed by the RIXS measurement [28, 29]. By comparing the RIXS spectra with the $\sigma_1(\omega)$ of our $(Sr_{1-x}Ca_x)_2IrO_4$ and $(Sr_{1-y}Ba_y)_2IrO_4$ thin-films, we have noticed that the energy of the spin-orbit exciton of $Sr_2IrO_4$ lies on top of the dip region in $\sigma_1(\omega)$, as marked by the red arrows in Fig. 3. As spin-orbit exciton is formed in the ground state of this system, electron-hole pairs partially fill up the $J_{eff} = 1/2$ band with electrons and the $J_{eff} = 3/2$ band with holes, respectively. Due to the partial occupation of electrons in the $J_{eff} = 1/2$ band by the spin-orbit exciton, the spectral weight of inter-site *d-d* transitions within the $J_{eff} = 1/2$ band are reduced with the dip structure appearing around 0.5 – 0.8 eV.

Multi-orbital Hubbard model calculations show consistent results with our proposed model of the spin-orbit exciton overlapping with an electronic continuum originating from inter-site $d$-$d$ transitions within the $J_{\text{eff}} = 1/2$ band. To calculate the spectral weight of the spin-orbit exciton spectra and $\sigma_1(\omega)$, we adopt the four-site cluster shown in Ref. [25]. For simplicity, we only take into account $t_{2g}$ orbitals and assume that the bond angle is 180°. The corresponding Hamiltonian is given by:

$$H = \sum_{i\delta\mu\nu\sigma} t_{\mu\nu}^{\delta} c_{i_\delta\mu\sigma}^\dagger c_{i\nu\sigma} + \sum_{i\mu\sigma} \Delta_{xy}\left(\delta_{\mu,xy} - \frac{1}{3}\right) c_{i\mu\sigma}^\dagger c_{i\mu\sigma} + \lambda \sum_{i\mu\nu\sigma\sigma'} (\vec{l}\cdot\vec{s})_{\mu\sigma,\nu\sigma'} c_{i\mu\sigma}^\dagger c_{i\nu\sigma'}$$

$$+ \frac{1}{2} \sum_{i\sigma\sigma'\mu\nu} U_{\mu\nu} c_{i\mu\sigma}^\dagger c_{i\nu\sigma'}^\dagger c_{i\nu\sigma'} c_{i\mu\sigma} + \frac{1}{2} \sum_{\substack{i\sigma\sigma' \\ \mu\neq\nu}} J_{\mu\nu} c_{i\mu\sigma}^\dagger c_{i\nu\sigma'}^\dagger c_{i\mu\sigma'} c_{i\nu\sigma}$$

$$+ \frac{1}{2} \sum_{\substack{i\sigma \\ \mu\neq\nu}} J_{\mu\nu}' c_{i\mu\sigma}^\dagger c_{i\mu\bar\sigma}^\dagger c_{i\nu\bar\sigma} c_{i\nu\sigma}, \quad (2)$$

where $c_{i\nu\sigma}^\dagger$ is the creation operator of an electron with $\nu$ orbital and $\sigma$ spin at the $i$-site and $i_\delta$ refers to the neighboring site of the $i$-th site whose displacement is $\vec{\delta}$. When $\vec{\delta} = \pm d\hat{x}$, where $d$ is the distance between neighboring sites, $t_{\mu\nu}^{\delta}$ has non-zero element ($t$) when $\mu = \nu = xy$ and $\mu = \nu = zx$. The second and third terms describe the local energy splitting due to the tetragonal distortion and spin-orbit coupling, respectively. The last three terms are the Hamiltonian of the local electron-electron correlation. We have parameterized correlation matrices as $U_{\mu\mu} = U$, $U_{\mu\neq\nu} = U - 2J_H$, and $J_{\mu\nu} = J_{\mu\nu}' = J_H$. We have set $\Delta_{xy} = 0.1$, $\lambda = 0.48$, $U = 1.86$, and $J_H = 0.5$ eV to be consistent with previous literature [25] and have fitted the experimental optical peak positions. For the $t_{2g}^5$ configuration, we consider all states in which the total number of holes is four. To solve Eq. 2, we employ the exact diagonalization method based on the Lanczos algorithm [30]. We calculate the ground state ($|\Psi_0\rangle$) and its energy ($E_0$) with the energy accuracy of $10^{-10}$ eV. To explore the distribution of the spin-orbit exciton, we calculate following projected excitation spectra:

$$\Lambda_2(\omega) = \sum_{nk}\langle\Psi_n|Q_k\rangle\langle Q_k|\Psi_n\rangle\delta(\omega - E_n + E_0) = -\frac{1}{\pi}\text{Im}\sum_k\langle Q_k|\frac{1}{\omega - H + E_0 + i\delta}|Q_k\rangle, \quad (3)$$

where $n$ and $|\Psi_n\rangle$ are the $n$-th eigenvalue and state, respectively. $|Q_k\rangle$ is the $k$-th orthonormal base state to span the subspace which consists of the spin-orbit exciton states with one $J_{\text{eff}} = 3/2$-hole in one site and one $J_{\text{eff}} = 1/2$-hole in other sites. Based on the continued fraction method [30], we solve Eq. (3). We have set $\delta = 0.03$ eV and perform 300 iteration steps. To calculate $\sigma_1(\omega)$, we use the following Kubo formula:

$$\sigma_1(\omega) = \frac{\pi v}{\omega}\sum_{n\neq 0}|\langle\Psi_n|\hat{J}_c|\Psi_0\rangle|^2\delta(\omega - E_n + E_0) = -v\text{Im}\sum_{n\neq 0}\frac{|\langle\Psi_n|\hat{J}_c|\Psi_0\rangle|^2}{(E_n - E_0)(\omega - E_n + E_0 + i\delta)}$$

$$= -v\text{Im}\left[\frac{1}{\omega + i\delta}\left(\langle\Psi_0|\hat{J}_c\frac{1}{H - E_0}\hat{J}_c|\Psi_0\rangle + \langle\Psi_0|\hat{J}_c\frac{1}{\omega - H + E_0 + i\delta}\hat{J}_c|\Psi_0\rangle\right)\right], \quad (4)$$

where $v$ is the volume per site and $\hat{J}_c$ is the current operator. We also exploit the continued fraction method with 400 iteration steps and $\delta = 0.1$ eV to solve Eq. (4). Note that increased $t$ from (Sr$_{1-x}$Ca$_x$)$_2$IrO$_4$ to (Sr$_{1-y}$Ba$_y$)$_2$IrO$_4$ leads to an increase in the bandwidth ($W$) of the system. We have used $t = 0.22$ eV for Sr$_2$IrO$_4$, which is a reasonable value for this iridate compound [31]. We have calculated $\Delta W\%$ as we substitute Sr with Ca or Ba using $\theta$ from Ref. [22] and Eq. (1). By considering the proportionality of $W$ with $t$, we have estimated the percentage change of $t$ ($\Delta t\%$). This calculation results in $t = 0.20$ eV and 0.23 eV for Ca and Ba substitution, respectively. Figures 5 (a) and (b) show the calculated spectral weight of the spin-orbit exciton and calculated $\sigma_1(\omega)$ for three different values of $t$ ( $t = 0.20$ eV for $x > 0$, $t = 0.22$ eV for $x = y = 0$, and $t =$ 0.23 eV for $y > 0$). Note that the spin-orbit exciton transition (Fig. 5 (a)) scales with the dip position in the experimental and calculated $\sigma_1(\omega)$ (Figs. 3 and 5 (b)).

Using our experimental and theoretical results, we suggest that the overall peak structure is solely from the inter-site $d$-$d$ transitions within the $J_{eff} = 1/2$ band and the dip structure around 0.5 – 0.8 eV in $\sigma_1(\omega)$ of layered iridates is a signature of the optically forbidden spin-orbit exciton. By decreasing $U_{eff}$ (increasing $t$) from (Sr$_{1-x}$Ca$_x$)$_2$IrO$_4$ to (Sr$_{1-y}$Ba$_y$)$_2$IrO$_4$, the overall $\sigma_1(\omega)$ at low energy (below 2 eV) red-shifts while the peak position of the spin orbit exciton does not change. The combination of the inter-site transitions within the $J_{eff} = 1/2$ band and the spin-orbit exciton creates a red-shifted two-peak feature in the low energy range of $\sigma_1(\omega)$ (Fig. 5 (b)). These theoretical results are consistent with the observed $\sigma_1(\omega)$, shown in Fig. 3.

Since the bandwidth of layered iridates can also change with the variation of temperature, it is worthwhile to compare our experimental results with the temperature-dependent $\sigma_1(\omega)$ from the bulk single crystal Sr$_2$IrO$_4$ [4, 5]. The temperature-dependent data indicate that the spectral weight of the $\alpha$ ($\beta$) peak increases (decreases) as the in-plane Ir-O-Ir bond angle increases. This behavior is consistent with our observation in thin-films in which the ratio of the spectral weight (SW$_\alpha$/SW$_\beta$) between $\alpha$ and $\beta$ optical transitions increases as it goes from (Sr$_{1-x}$Ca$_x$)$_2$IrO$_4$ to (Sr$_{1-y}$Ba$_y$)$_2$IrO$_4$ films (Fig. 4 (b)). Further, as the bond angle increases due to thermal expansion, the $\alpha$ peak shifts to lower energy which is consistent with our A-site dependence data as well. However, the red-shift of the $\beta$ peak (by increasing temperature) is less visible in temperature-dependent data [4, 5]. Based on our interpretation of $\sigma_1(\omega)$, as the temperature increases ($U_{eff} = U/W$ decreases), the overall spectra shifts to lower energy. However, the spin orbit exciton may also be temperature-dependent and move to higher energy, which acts to reduce the resultant red-shift in the $\beta$ peak position. It is necessary to investigate the temperature-dependence of the spin-orbit exciton to fully understand this picture. As shown schematically in Fig. 5 (c), the spin-orbit exciton and the inter-site optical transition of the $J_{\text{eff}} = 1/2$ state strongly overlap, which creates a dip structure in $\sigma_1(\omega)$, resulting in the two-peak structure observed in $\sigma_1(\omega)$ of layered iridates system.

The $\sigma_1(\omega)$ of bandwidth controlled $(Sr_{1-x}Ca_x)_2IrO_4$ and $(Sr_{1-y}Ba_y)_2IrO_4$ films provide indirect evidence regarding the nature of the spin-orbit exciton. Hence, we suggest advanced spectroscopic characterizations such as resonant inelastic x-ray scattering experiments on this system to confirm this picture. Moreover, in order to fully understand this $(Sr_{1-x}A_x)_2IrO_4$ thin-film (A: Ca, Ba) system, it is important to obtain the local structural information such as octahedral rotation and tilting as a future study. We have assumed that doping with smaller (larger) A-site ions increases (decreases) rotation and tilting in the system, which is consistent with previous work [19]. However, decrease in the in-plane rotation by the substitution of Ca for Sr has been also suggested recently [18, 32].

Our approach of studying bandwidth controlled epitaxial thin-films provides a useful way to unveil controversial issues in strongly correlated materials. By applying chemical pressure beyond the solubility limit, we can control the lattice parameters, Ir-O-Ir bond angle, electronic hopping, and electronic correlation effects in the system. Recently, it has been reported that even a slight increase in the Ir-O-Ir bond angle can create a huge increase in the electronic hopping of compounds like $Sr_3Ir_2O_7$. This can cause a drastic decrease in the resistivity of the system and create a metallic state or possibly even a superconducting state [33, 34]. Hence, studying iridates under factors like chemical pressure can help us explore these systems and potentially uncover the novel properties that are theoretically predicted.

## V.     SUMMARY

We have synthesized and investigated epitaxial thin-films of $(Sr_{1-x}Ca_x)_2IrO_4$ and $(Sr_{1-y}Ba_y)_2IrO_4$ ($x$ and $y$ = 0 - 0.375), which effectively act to tune the A-site ionic radius of the layered iridate system and subsequently its bandwidth. Using a systematic study of the A-site dependence on $\sigma_1(\omega)$ with respect to the change in the bandwidth, we have observed red-shifted optical peak positions in low energies from $(Sr_{1-x}Ca_x)_2IrO_4$ to $(Sr_{1-y}Ba_y)_2IrO_4$. This unexpected observation cannot be explained using the conventional $J_{eff} = 1/2$ and $J_{eff} = 3/2$ band picture. Our experimental observations are consistent with theoretical results using multi-orbital Hubbard model calculations that suggest the spin-orbit exciton and the inter-site optical transition of the $J_{eff} = 1/2$ state strongly overlap due to a Fano-like resonance. This imposes a dip structure in $\sigma_1(\omega)$ and creates a two peak structure in the spectra. The optical peak positions at low energy red-shift as $U_{eff}$ decreases from $(Sr_{1-x}Ca_x)_2IrO_4$ to $(Sr_{1-y}Ba_y)_2IrO_4$. Our results confirm that controlling the bandwidth of layered iridates can help resolve the controversial issues in understanding the electronic structure of this system. Moreover, the epitaxial growth of thin-films under chemical pressure is a viable technique to expand the scope of materials with competing interactions and provides a platform for investigating the existing arena for novel phenomena in strongly correlated materials.

## VI.   ACKNOWLEDGMENTS

We thank C. H. Sohn for useful discussions and valuable comments. We acknowledge the support of National Science Foundation grants DMR-1454200 for thin-film synthesis and characterizations, DMR-1265162 and DMR-1712101 for target synthesis, and DMR-1262261 for infrared spectroscopy. B.H.K. acknowledges the support from the RIKEN iTHES Project for the numerical calculations.

**FIGURE CAPTIONS**

**FIG. 1.** (a) X-ray $\theta$-$2\theta$ scans of the epitaxial $(Sr_{1-x}A_x)_2IrO_4$ (A: Ca, Ba) thin-films grown on STO substrates, where only the (00$l$)-diffraction peaks of the films ($l$ = 4, 8, 12, 16, 24) are visible. (b) The enlarged scans near (00$\underline{12}$) reflections of the films and the (002) reflections of the substrates. The peaks from the substrates are labeled with the filled diamond (♦) symbols.

**FIG. 2.** (a) Reciprocal space map around the (103) reflection of the STO substrates with the (11$\underline{18}$) reflection of the $(Sr_{1-x}Ca_x)_2IrO_4$ and $(Sr_{1-y}Ba_y)_2IrO_4$ films. (b) The in-plane (left axis) and out of plane (right axis) lattice parameters of the $(Sr_{1-x}Ca_x)_2IrO_4$ and $(Sr_{1-y}Ba_y)_2IrO_4$ thin-films obtained from the reciprocal space maps and x-ray diffraction scans, respectively, as a function of average A-site ionic radius in the $(Sr_{1-x}A_x)_2IrO_4$ thin-films (A: Ca, Ba). The solid red and blue squares represent the in-plane and out of plane lattice parameters, respectively. (c) Normalized resistivity versus temperature of the $(Sr_{1-x}Ca_x)_2IrO_4$ and $(Sr_{1-y}Ba_y)_2IrO_4$ thin-films on STO substrates which indicates that all the films are insulators.

**FIG. 3.** Optical conductivity spectra ($\sigma_1(\omega)$) of the $(Sr_{1-x}Ca_x)_2IrO_4$ and $(Sr_{1-y}Ba_y)_2IrO_4$ thin-films at room temperature with the minimal set of Lorentz-oscillators. The experimental spectra are shown by thick olive, cyan, blue, magenta, and orange curves from $(Sr_{0.75}Ca_{0.25})_2IrO_4$ to $(Sr_{0.75}Ba_{0.25})_2IrO_4$, respectively. The fit oscillators are shown by thin green, red, blue, orange and gray curves. The black curves show the total spectra of the fit, which matches with the experimental spectra. Note that the $\beta$ peak position shows a significant red-shift as it goes from $(Sr_{1-x}Ca_x)_2IrO_4$ to $(Sr_{1-y}Ba_y)_2IrO_4$. The spectra are shifted vertically for clarity.

**FIG. 4.** (a) The $\alpha$ and $\beta$ peak positions ($\omega_\alpha$ and $\omega_\beta$) as a function of average A-site ionic radius for the $(Sr_{1-x}A_x)_2IrO_4$ thin-films (A: Ca, Ba), obtained from Lorentz oscillator fitting. (b) The ratio of the spectral weight of the $\alpha$ to $\beta$ peaks versus average A-site ionic radius from $(Sr_{1-x}Ca_x)_2IrO_4$ to $(Sr_{1-y}Ba_y)_2IrO_4$.

**FIG. 5.** (a) Calculated spectral weight of spin-orbit exciton and (b) $\sigma_1(\omega)$ at different values of the electronic hopping integral ($t$). The spectra in (b) are shifted vertically for clarity. (c) Schematic diagram representing the Fano-like resonance between the inter-site optical transitions of the $J_{eff} = 1/2$ state and the spin-orbit exciton, which creates a two peak-like structure in the final optical spectrum. The red and blue arrows are electrons with up and down spins in the $J_{eff} = 3/2$ and $J_{eff} = 1/2$ states, indicating the inter-site optical transitions and the on-site spin-orbit exciton.

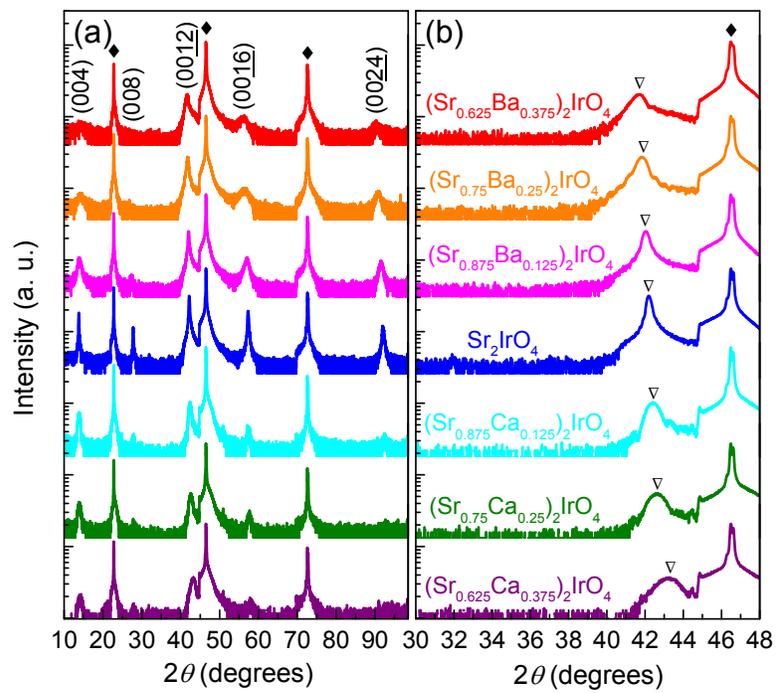

Fig. 1
M. Souri *et al.*

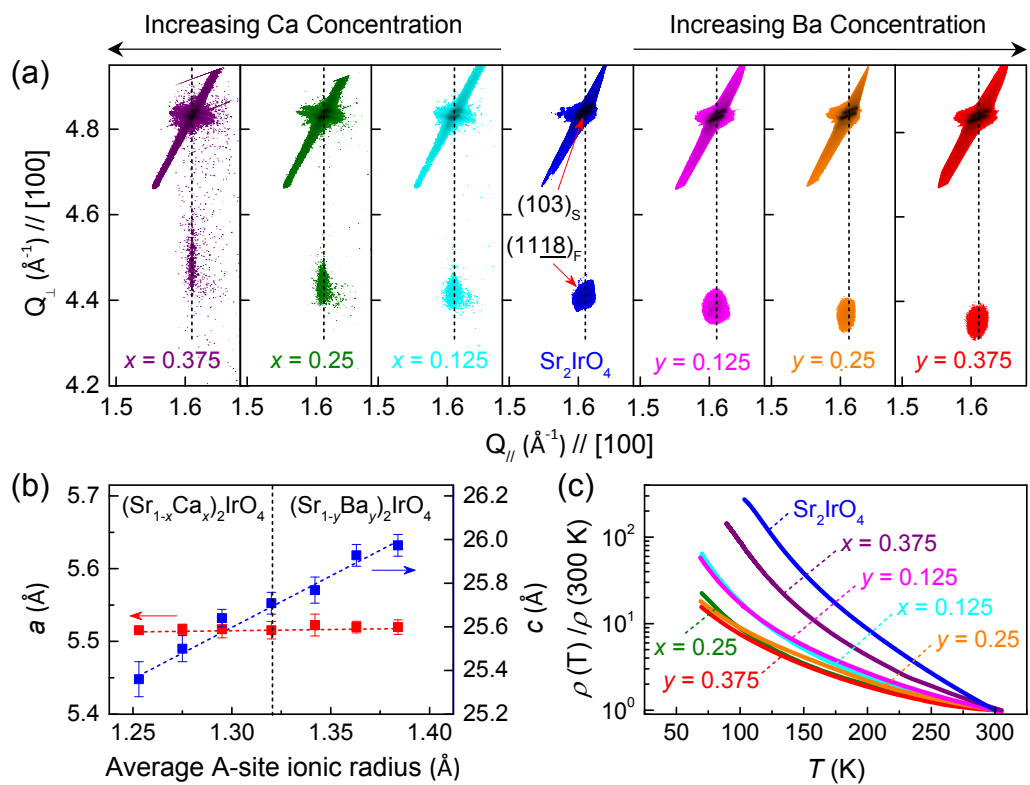

Fig. 2
M. Souri et al.

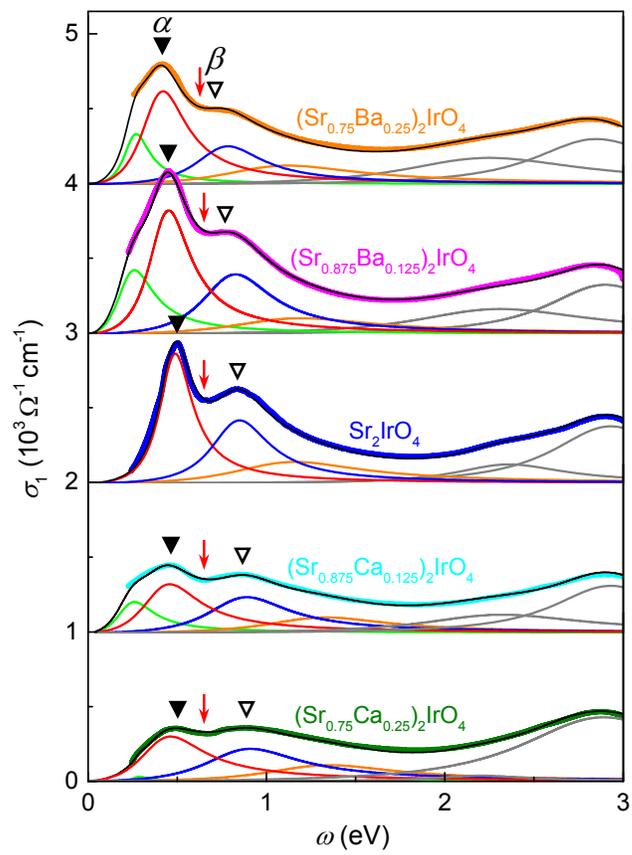

Fig. 3
M. Souri *et al.*

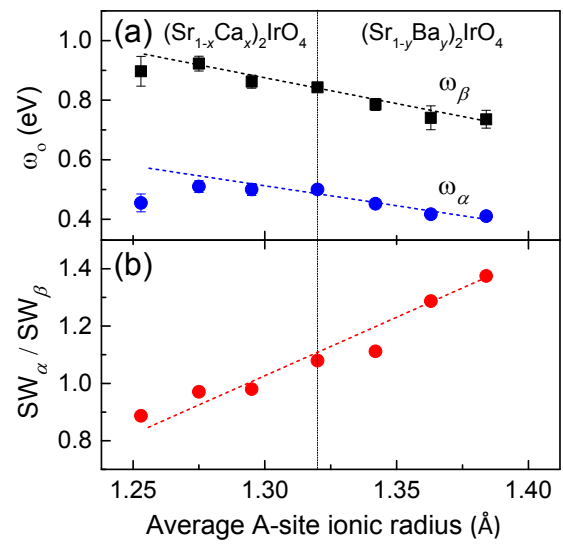

Fig. 4
M.Souri *et al.*

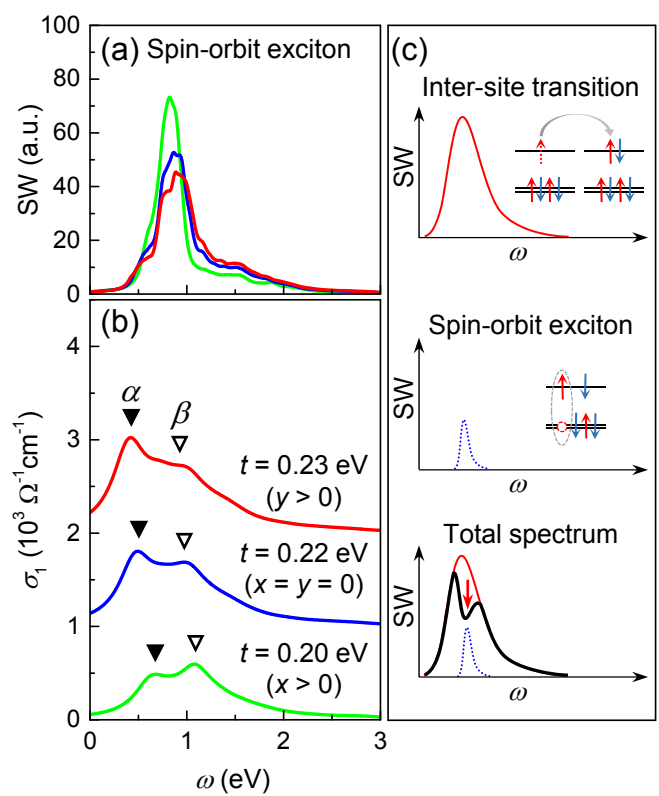

Fig. 5
M. Souri *et al.*